# Seed-Driven Stepwise Crystallization (SDSC) for Growing Rutile $GeO_2$ Films via MOCVD

Imteaz Rahaman[1], Botong Li[1], Bobby Duersch[2], Hunter D. Ellis[1], and Kai Fu[1,*]

[1]*Department of Electrical and Computer Engineering, The University of Utah, Salt Lake City, UT 84112, USA*

[2]*Electron Microscopy and Surface Analysis Laboratory, The University of Utah, Salt Lake City, UT 84112, USA*

**Abstract**

Germanium dioxide (r-$GeO_2$) is an emerging ultrawide bandgap (UWBG) semiconductor with significant potential for power electronics, thanks to its ambipolar doping capability. However, phase segregation during metal-organic chemical vapor deposition (MOCVD) on substrates like r-$TiO_2$ has posed a significant barrier to achieving high-quality films. Conventional optimization of growth parameters has been found so far not very insufficient in film coverage and film quality. To address this, a seed-driven stepwise crystallization (SDSC) growth approach was employed in this study, featuring multiple sequential deposition steps on a pre-templated substrate enriched with r-$GeO_2$ seeds. The process began with an initial 180-minute deposition to establish r-$GeO_2$ nucleation seeds, followed by a sequence of shorter deposition steps (90, 60, 60, 60, 60, and 60 minutes). This stepwise growth strategy progressively increased the crystalline coverage to 57.4%, 77.49%, 79.73%, 93.27%, 99.17%, and ultimately 100%. Concurrently, the crystalline quality improved substantially, evidenced by a ~30% reduction in the Full Width at Half Maximum (FWHM) of X-ray diffraction rocking curves. These findings demonstrate the potential of the SDSC approach for overcoming phase segregation and achieving high-quality, large-area r-$GeO_2$ films.

* Author to whom correspondence should be addressed. Electronic email: kai.fu@utah.edu

## Introduction

Ultrawide-bandgap (UWBG) semiconductors are becoming indispensable for next-generation power electronics and optoelectronic devices due to their exceptional electrical, thermal, and optical properties. Materials like β-$Ga_2O_3$, AlGaN, and diamond exhibit high breakdown fields and superior power-handling capabilities, making them ideal for applications such as high-power transistors, UV detectors, and gas sensors.[1–3] However, despite their promise, each material faces intrinsic challenges: β-$Ga_2O_3$ suffers from inefficient heat dissipation and a lack of p-type doping;[4–6] AlGaN experiences degraded acceptor activation and hole mobility with increasing Al content;[7,8] diamond production is hindered by limited wafer availability and doping difficulties.[9,10] Rutile $GeO_2$ (r-$GeO_2$) has recently emerged as a promising UWBG semiconductor, exhibiting a bandgap of 4.44–4.68 eV[11–13], electron mobility of 244 $cm^2/V\cdot s$ ⊥C and 377 $cm^2/V\cdot s$ ∥C[14], and a Baliga figure of merit[15] of 27,000–35,000 $\times 10^6$ $V^2$ $\Omega^{-1}$ $cm^{-2}$ surpassing that of β-$Ga_2O_3$. Its thermal conductivity, measured at 37 W/m·K (*a*-axis) and 58 W/m·K (*c*-axis), is nearly double that of β-$Ga_2O_3$.[16] Moreover, it is theoretically predicted to support ambipolar doping with promising hole mobility [27 $cm^2$/Vs ($\perp\vec{C}$)) and 29 ($\parallel\vec{C}$)], paving the way for p-n junctions and flexible design of various power devices.[12,14,17] Furthermore, the rutile phase of $GeO_2$ is water-insoluble, unlike its α-quartz counterpart.[18] It is also likely to demonstrate potential for bandgap tuning from 3.03 to 8.67 eV through alloying with other rutile oxides, such as r-$SnO_2$, r-$TiO_2$, and r-$SiO_2$, similar to III-nitrides and $(Al_xGa_{1-x})_2O_3$.[19–23] Furthermore, the rutile $TiO_2$ has a small lattice mismatch with r-$GeO_2$,[24,25] making it a practical substrate for r-$GeO_2$ epitaxy. However, achieving high-quality and wafer-scale r-$GeO_2$ films remains challenging due to the competition between other phases, such as α-quartz and amorphous $GeO_2$. Current epitaxial growth techniques, including molecular beam epitaxy (MBE),[26] mist chemical vapor deposition (CVD),[24] metal-organic chemical vapor deposition (MOCVD)[27,28] sputtering technique[29–31] and pulsed laser deposition (PLD),[32–34] have reported limited success in stabilizing the rutile phase. It is worth noting that Rahaman *et al.* demonstrated via MOCVD that while a 90-minute deposition enhanced film quality, a 180-minute deposition increased coverage of rutile $GeO_2$ but reduced quality.[27] Consequently, achieving full crystalline coverage while preserving good crystalline quality remains challenging during continuous deposition, regardless of whether the deposition time is short or long, due to phase segregation and defect formation.

In this work, a seed-driven stepwise crystallization (SDSC) approach has been implemented to overcome the epitaxy challenges by MOCVD. Through utilizing multiple deposition steps on the seeded template, the approach leverages nucleation seeds formed during the earlier stages of growth to enable full crystalline coverage of r-$GeO_2$ films on r-$TiO_2$ (001) substrates. The films were comprehensively characterized using techniques such as High-Resolution X-ray diffraction (HR-XRD), scanning electron microscopy (SEM), and reciprocal space mapping (RSM) to evaluate their crystal quality, morphology, and stress.

## Experimental details

Figure 1 illustrates the experimental process and outlines the steps involved in the Seed-Driven Stepwise Crystallization (SDSC) method. The experiment was designed based on the following strategies and key findings, as shown in Fig. 1(a): [1]The coexistence of different $GeO_2$ phases (rutile, quartz, and amorphous) suggests competition between these phases. [2] Continuous growth can lead to phase separation, with longer growth periods resulting in lower quality. For example, a 180-minute growth shows poorer quality than a 90-minute growth.[27] [3] Optimizing growth conditions to favor the rutile phase promotes its epitaxy, allowing it to prevail over other phases when grown on a rutile-seeded substrate. [4] Repeating the process [3] will eventually cause the rutile phase to cover the entire substrate surface.

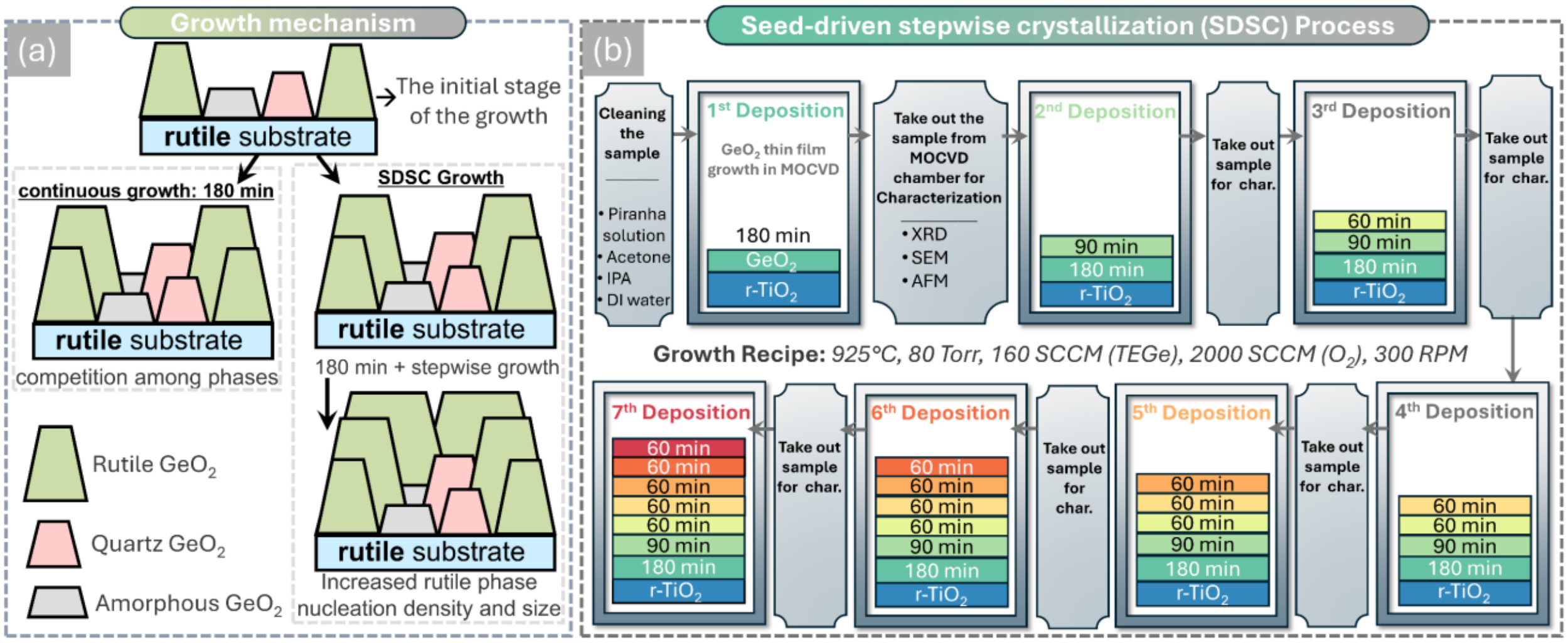

**FIG. 1.** (a) Strategies and mechanisms considered for the experiments. (b) SDSC growth process for growing r-$GeO_2$ films on r-$TiO_2$ (001) substrates repeating the growth-seeding-characterization cycles with growth durations of 180 min for the 1$^{st}$ one, 90 min for the 2$^{nd}$ one, and 60 min for depositions from the 3$^{rd}$ one.

As shown in Fig. 1(b), prior to loading the r-$TiO_2$ (001) substrate into the MOCVD chamber, a thorough cleaning procedure was performed to ensure a contaminant-free surface. This process began with treatment using a piranha solution ($H_2SO_4$:$H_2O_2$ = 3:1) to remove organic contaminants, followed by sequential rinsing with acetone, isopropanol, and deionized water. Once cleaned, the r-$TiO_2$ substrate was placed into the MOCVD chamber. The depositions were carried out in a customized MOCVD reactor (manufactured by Agnitron Technology), with an active chamber volume of 967.72 $cm^3$, a chamber diameter of 95 mm, and a distance from the top of the sample holder to the showerhead of 136.5 mm. The deposition process was conducted at a temperature of 925°C under a stable chamber pressure of 80 Torr. Tetraethyl germane (TEGe) and pure oxygen ($O_2$) were utilized as precursors, with argon (Ar) functioning as both the carrier and shroud gas. The flow rates were 2000 SCCM for oxygen and 160 SCCM for the TEGe precursor while maintaining a susceptor rotation speed of 300 RPM for uniform deposition. The SDSC process began with an initial 180-minute deposition, followed by characterization of the film. Simultaneously, the first rutile-seeded substrate was formed. This growth-seeding-characterization cycle was repeated a total of seven times, with varying deposition times. The growth duration was reduced to 90 minutes for the 2nd deposition and 60 minutes for all subsequent depositions starting from the 3rd deposition, to investigate the evolution of increased seeding and coverage of r-$GeO_2$ films. During each deposition, the sample was cooled down to room temperature at a rate of ~30°C/min, with no additional temperature holds before or after each deposition. Table 1 provides a detailed summary of the sample IDs and the corresponding growth durations for each step.

Table 1. Summary of sample IDs, corresponding growth steps, and their respective growth durations during the SDSC process.

| **Sample ID** | **Sequence** | **Growth duration (min)** |
|---|---|---|
| T-22(0) | Step-1 | 180 |
| T-22(1) | Step-2 | 90 |
| T-22(2) | Step-3 | 60 |
| T-22(3) | Step-4 | 60 |
| T-22(4) | Step-5 | 60 |
| T-22(5) | Step-6 | 60 |
| T-22(6) | Step-7 | 60 |

The structural characterization of the $GeO_2$ thin films was performed using a Bruker D8 DISCOVER high-resolution XRD equipped with a Cu $K\alpha_1$ source ($\lambda = 1.5406$ Å), a triple-bounce channel-cut monochromator, and an Eiger R 250K detector. Surface morphology analysis was conducted using a Micron scale EDS Quanta 600F environmental SEM and a Bruker Dimension ICON AFM. To quantify the coverage of the rutile phase, ImageJ software is used to measure the area ratio between the rutile and non-rutile regions.

## Results and analysis

Figure 2 illustrates the SEM images of the progressive evolution of r-$GeO_2$ coverage and morphology on the r-$TiO_2$ (001) substrate during the SDSC process. In the initial deposition stage (Step 1), isolated nucleation seeds of rutile r-$GeO_2$ (bright regions) are distinctly visible amidst significant uncovered substrate areas, representing the early stage of sparse island growth. As the process advances to the second deposition (Step 2), the nucleation density and island size noticeably increase and a distinct contrast emerges between the rutile $GeO_2$ phase (bright regions)[35] and the quartz/amorphous $GeO_2$ phase (dark regions)[35].

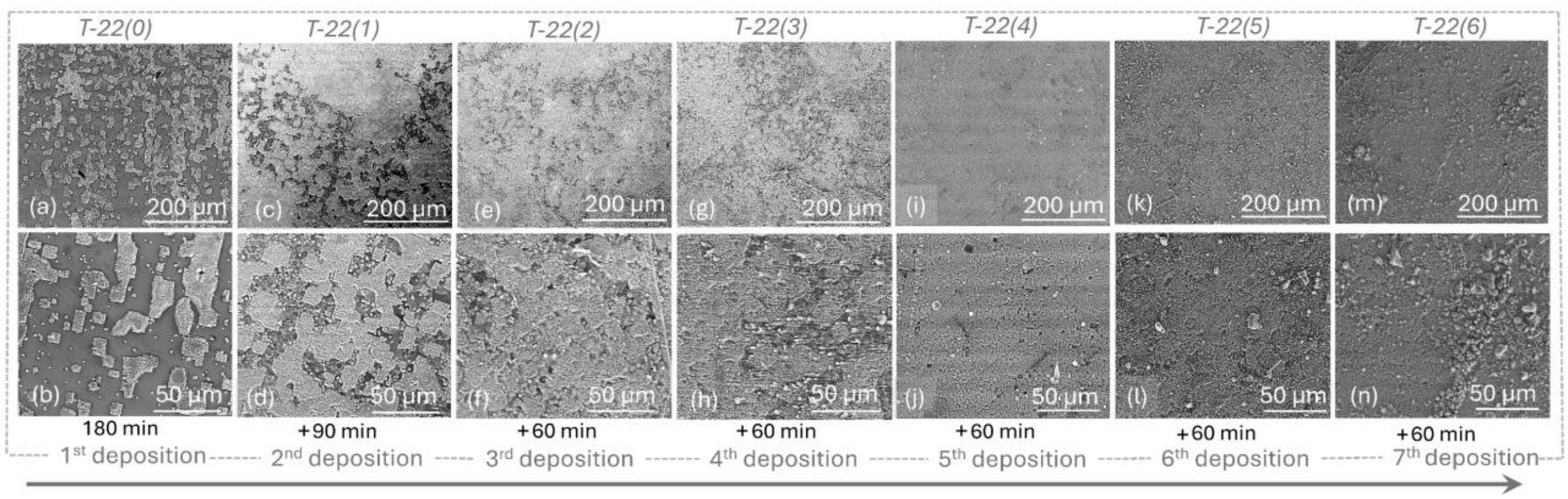

**FIG. 2.** SEM images showing the stepwise evolution of r-$GeO_2$ coverage on r-$TiO_2$ (001) substrate during the SDSC growth process. The progression highlights the increase in nucleation density and the dominance of rutile crystals through lateral growth and coalescence.

This multiple-phase coexistence arises from the thermodynamic competition among rutile $GeO_2$ and other non-rutile phases, including quartz-like and amorphous structures. During the intermediate stages (Steps 3 and 4), the rutile phase progressively dominates as nucleation sites expand laterally and vertically. This growth promotes the merging of adjacent crystalline domains, increasing coverage and leading to enhanced coalescence of rutile crystals. The stepwise growth strategy enhances adatom mobility and lateral diffusion, allowing nucleation seeds to coalesce into larger domains while minimizing competing phase formation.[36–38] The SDSC process enhances adatom mobility and lateral diffusion by introducing controlled growth interruptions between deposition steps. In contrast, the continuous growth (180-min growth) results in a coexistence of various phases (rutile, quartz, and amorphous). The SDSC approach forms rutile $GeO_2$ seed islands by the 1st deposition on the rutile $TiO_2$ substrate, which act as nucleation sites promoting lateral expansion rather than random nucleation in the following deposition. The shorter growth duration and cooling periods between deposition steps help stabilize the rutile phase seeds and prevent the seeding of other phases with lower energies than the rutile phase, even though the substrate is rutile phase. Specifically, the pathways available for the relaxation of adatoms depend on the nature of the bond between the edge atoms and the substrate, as well as the interaction with neighboring atoms on the existing rutile islands.[37] The stepwise process may also enhance the lateral diffusion of adatoms, facilitated by mechanisms such as terrace diffusion and step-edge attachment,[37] which encourages the coalescence of rutile islands and eventual full surface coverage. As confirmed by SEM images, the rutile nuclei expand laterally with each deposition step. Moreover, the coalescence of r-$GeO_2$ islands occurs to minimize the total surface energy of the system, as smaller islands tend to have higher energy due to their larger surface-to-volume ratios. This thermodynamic driving force leads to the merging of nucleation sites.[39,40] The amorphous, and quartz phases are expected to be buried, as predicted by thermodynamic stability models.[26,41] In the later stages (Steps 5 through 7), the r-$GeO_2$ crystals become increasingly prominent as square-shaped structures, with the amorphous and quartz phases largely suppressed. After the 7th deposition, the final film is entirely covered by the rutile phase, with other phases situated beneath the rutile layer. In contrast, continuous growth leads to phase competition between rutile and other phases, resulting in nonuniform morphology and reduced material quality.

Figure 3 illustrates the corresponding evolution of surface morphology. Initially, the growth mode is characterized by isolated island nucleation (Step 1), where individual crystallites form.

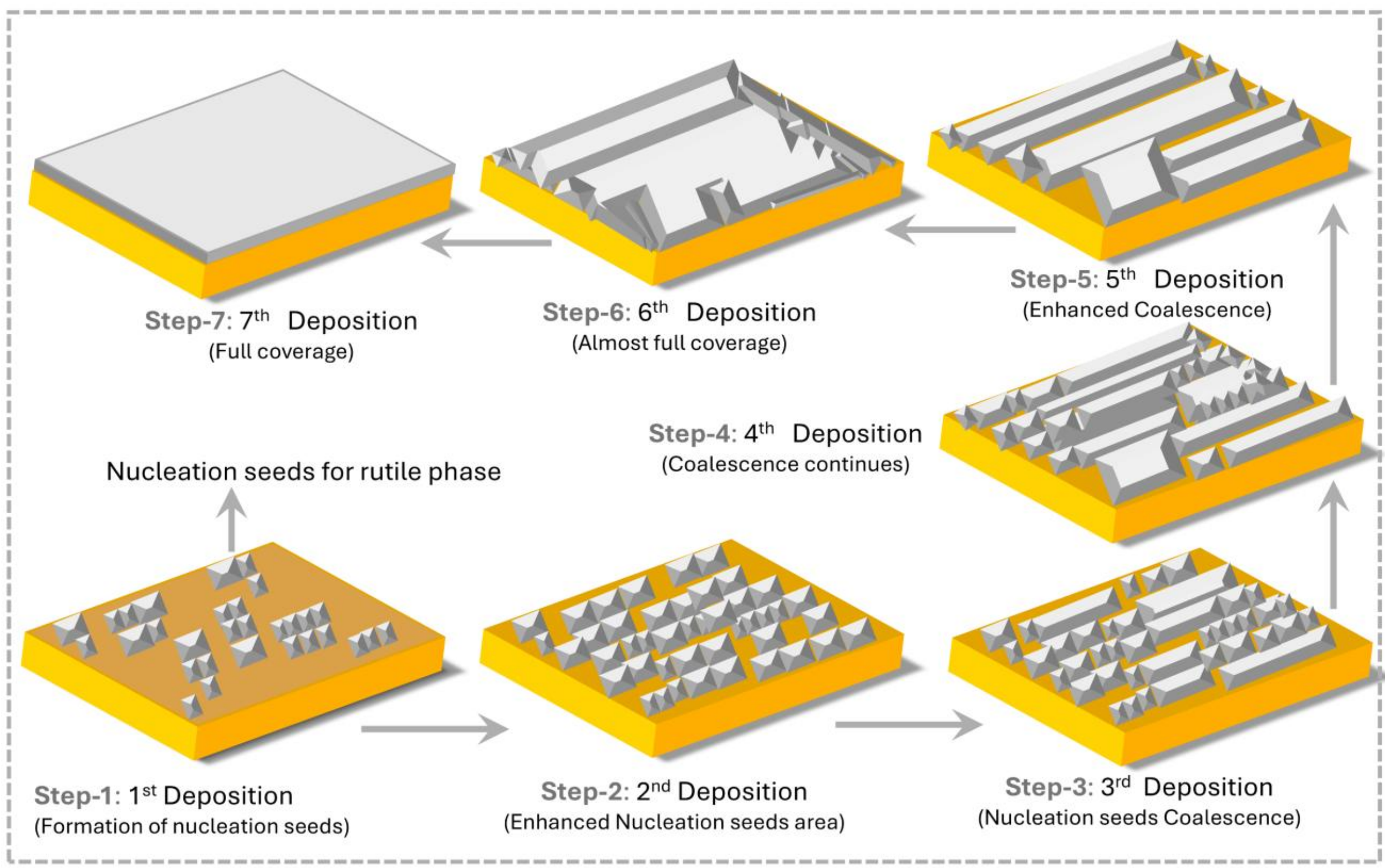

**FIG. 3.** Stepwise growth process of r-$GeO_2$ films using the SDSC method, showing nucleation seed formation (Step 1), enhanced nucleation area (Step 2), seed coalescence (Steps 3–5), and progression to full surface coverage (Steps 6–7).

Over subsequent steps, lateral growth and vertical expansion of these islands occur, driven by surface diffusion and adatom mobility, which enhance the merging of adjacent crystalline domains.[37,42] During the intermediate stages (Steps 3–5), coalescence dominates as the crystalline patterns coalesce into larger structures, minimizing surface energy and increasing coverage.[36,39]

After Step 6 and Step 7, nearly full crystalline coverage is achieved, resulting in a fully continuous r-$GeO_2$ film.

Figure 4(a) presents the XRD $2\theta$-$\omega$ scan, where the prominent peak corresponding to the r-$GeO_2$ (002) plane, becomes increasingly intense with each deposition step, indicating a steady improvement in crystalline coverage. In contrast, peaks associated with other $GeO_2$ phases, such as cubic or hexagonal polymorphs, remain weak and diminish further in later steps, highlighting the selective stabilization and dominance of the rutile phase. This structural enhancement aligns well with the SEM observations, where the amorphous and quartz phases gradually become encapsulated beneath the growing rutile domains. Figure 4(b) shows the rocking curve (ω-scan) measurements, revealing a progressive reduction in the FWHM ($\Delta\omega$) from 0.37° after the first step to 0.259° after the final step. The consistent decrease in FWHM signifies a reduction in structural defects and improved crystalline uniformity. The decrease in the FWHM observed is primarily due to an increase in crystallite/domain size as the deposition progresses, resulting from the growth and coalescence of the rutile $GeO_2$ islands during the stepwise growth process. Complementing this, Figure 4(c) plots the trends of FWHM (yellow) and rutile crystal coverage (gray), showing a clear correlation between decreasing FWHM and increasing coverage. Starting from ~45% in the initial step, the coverage steadily improves to ~100% in the final step.

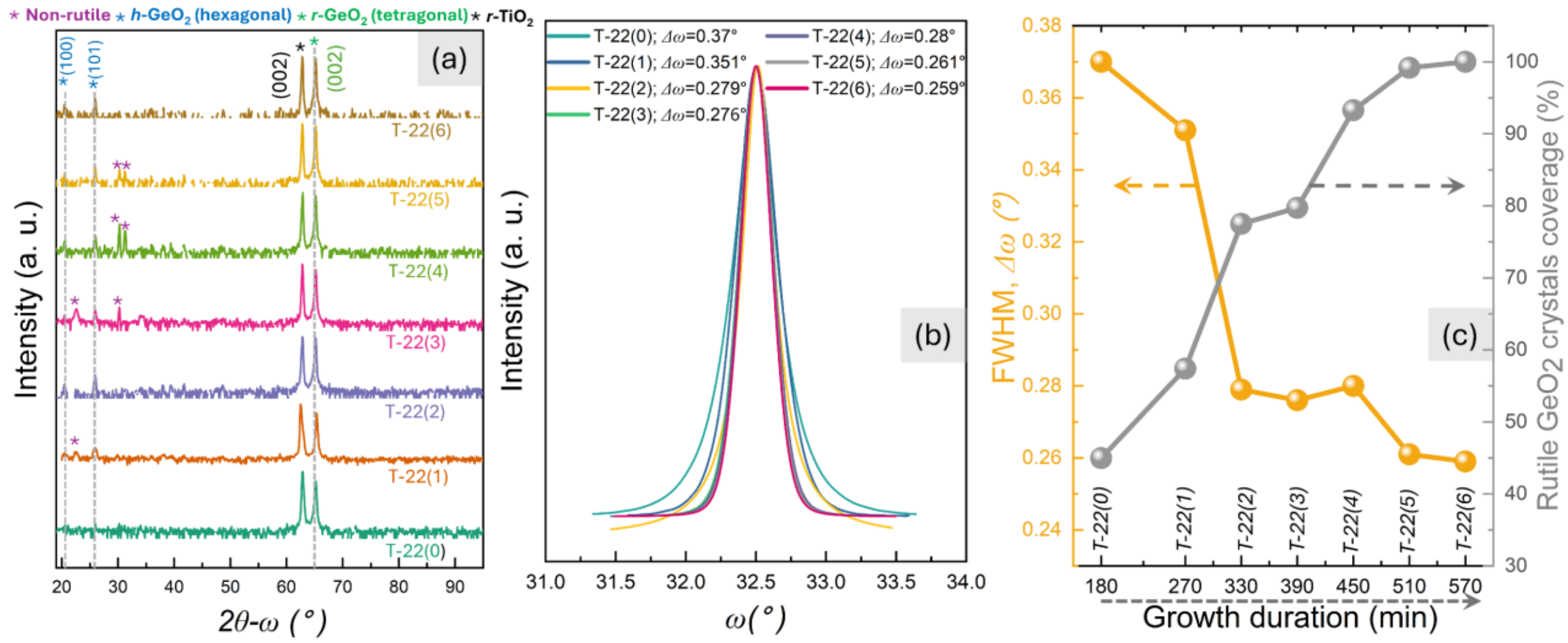

**FIG. 4.** Structural and morphological analysis of r-$GeO_2$ films during the SDSC growth process. (a) XRD *2θ-ω* scan showing the dominance of the rutile r-$GeO_2$ (002) peak. (b) Rocking curve (*ω-scan*) measurements demonstrating progressive improvement in crystalline quality with reduced FWHM (*Δω*). (c) Trends of FWHM (yellow) and r-$GeO_2$ crystal coverage (gray), highlighting simultaneous enhancement in crystalline quality and surface coverage through stepwise growth.

Figure 5 presents RSMs of symmetrical (002) and asymmetrical (022) reflections for the initial [T-22(0)] and final [T-22(6)] deposition steps of r-$GeO_2$ films grown on the r-$TiO_2$ (001) substrate. In the symmetrical RSMs [Figures 5(a) and 5(c)), the $GeO_2$ (002) peak aligns closely with the $TiO_2$ (002) substrate peak along the $q\|$-direction across all steps, indicating minimal tilt of the *c*-axis and maintaining crystallographic orientation relative to the substrate. The asymmetrical RSMs displayed in Figures 5(b) and 5(d) show the (022) reflections of $GeO_2$ and $TiO_2$. These reflections are characterized by $K\alpha_1$ splitting and minimal broadening in the $q\perp$-direction, which indicates the high structural quality of the films. Compared to T-22(0), the reflections in T-22(6) show less broadening in both the $q\|$ and $q\perp$-directions, reflecting improved crystallographic alignment and reduced mosaicity, respectively,[43] thanks to the stepwise deposition process. However, the $q\|$ and $q\perp$ values of the film reflections deviate from those of the substrate reflections, as highlighted by the triangular markers generated by Leptos 7.14 software for strain analysis. This shift indicates the presence of strain in films.

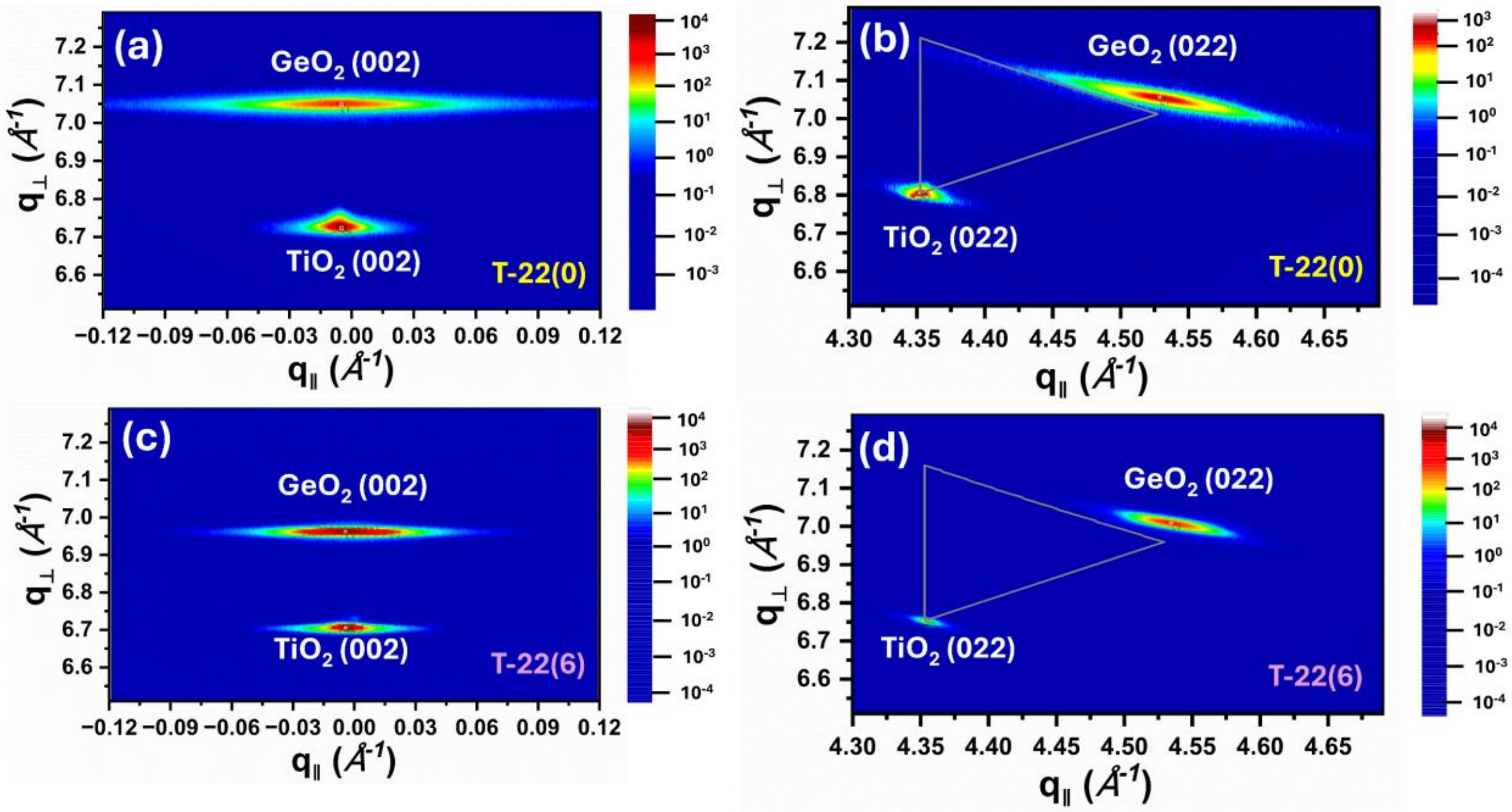

**FIG. 5.** Reciprocal space maps of r-$GeO_2$ films on r-$TiO_2$ substrates. Panels (a), and (b) show symmetric (002) reflections, while panels (c) and (d) display asymmetric (022) reflections for samples T-22(0) and T-22(6), respectively.

Quantitative analysis of the strain data from RSM plots reveals that the in-plane (lateral) strain increased from 0.757% for T-22(0) to 0.978% for T-22(6), while the out-of-plane (normal) strain also increased from 0.467% for T-22(0) to 0.603% for T-22(6). These results indicate that while crystalline quality improves, films experience a slight increase in strain as growth progresses. This behavior may arise due to the gradual coalescence of nucleation seeds into a continuous film, which introduces residual strain as the film grows.[40]

Figure 6 shows the AFM surface morphology of initial deposition T-22(0) and final deposition T-22(6). The RMS roughness of the r-$GeO_2$ films is non-uniform across both the initial and final deposition steps. In the first step (T-22(0)), the roughness varies significantly, with an average RMS roughness of ~121.39 nm. After the final step (T-22(6)), the roughness remains non-uniform, with a measured average of 140.4 nm. While the SDSC method successfully enhances crystalline growth, further optimization is needed to reduce the overall roughness.

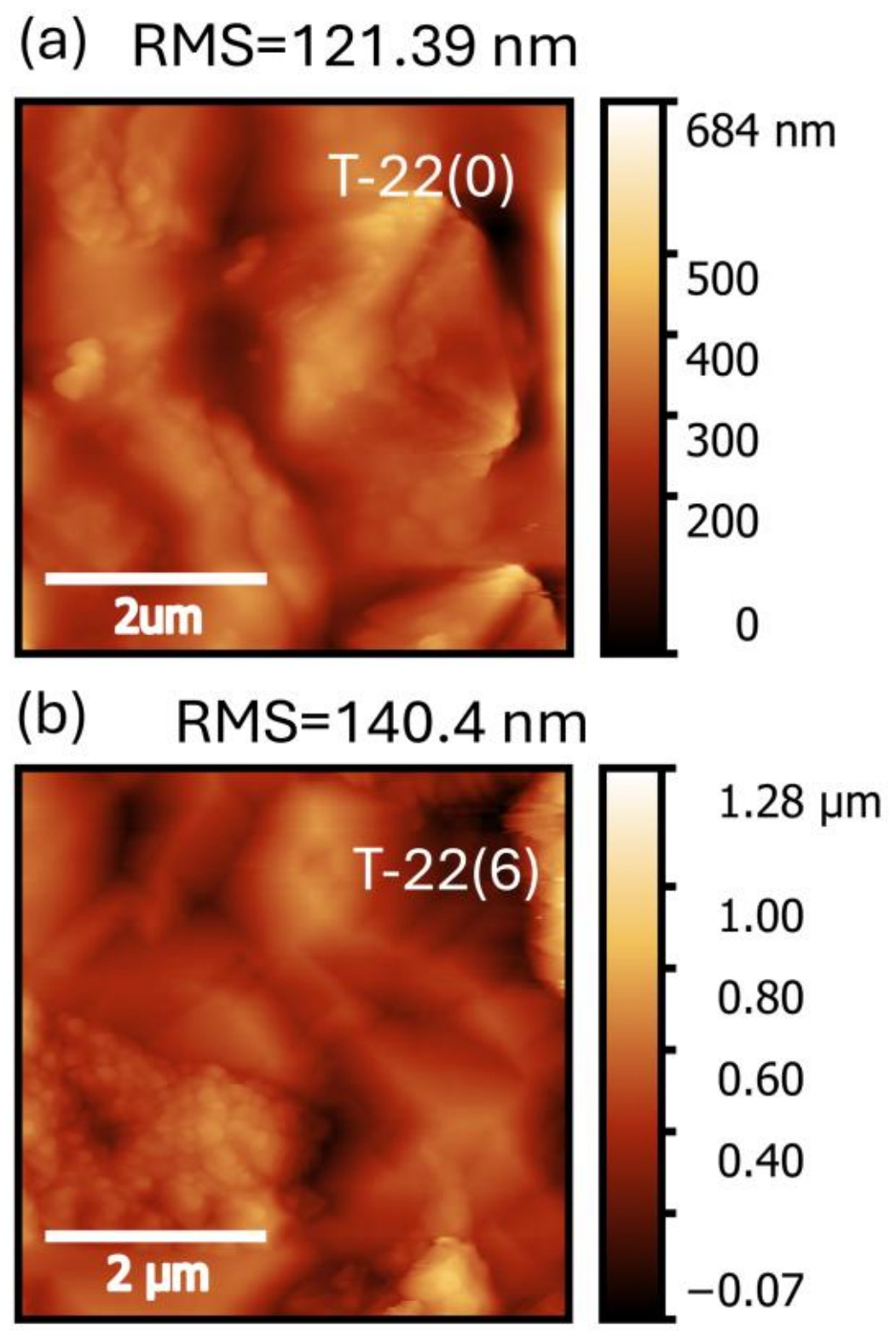

**FIG. 6.** Atomic Force Microscopy (AFM) surface morphology of the $GeO_2$ films after (a) 1$^{st}$ deposition (T-22(0)), and (b) final deposition (T-22(6)).

Growing high-quality films with low residual strain and smooth surfaces is essential in MOCVD. To reduce the residual strain and surface roughness in rutile $GeO_2$ films, several strategies can be considered. One effective approach is substrate optimization, which involves selecting a substrate that closely matches the lattice of the film material to minimize strain. Additionally, ensuring the substrate surface is clean and atomically flat is crucial, often requiring thorough cleaning and sometimes annealing procedures. For certain materials, slightly offcut substrates can encourage step-flow growth, further helping to reduce surface roughness. Another key strategy for reducing strain involves the use of strain relaxation layers or buffer layers. These layers, which have intermediate lattice constants, can gradually reduce strain in the film. Moreover, growing superlattices with alternating layers under tensile and compressive strain may help balance the overall strain across the film, improving both film quality and surface morphology.

## Conclusion

In conclusion, a successful growth of high-coverage, crystalline rutile $GeO_2$ films on r-$TiO_2$ (001) substrates has been demonstrated using the SDSC method. Through sequential deposition steps, we achieved a significant improvement in both crystalline quality and surface coverage. The XRD and RSM analyses confirmed the dominance of the rutile phase while the rocking curve (FWHM) revealed a systematic reduction of ~30% in FWHM from 0.37° after the initial stage to 0.259° after the final step, indicating enhanced crystalline quality. SEM observations demonstrated the progressive evolution of crystalline square patterns, with coverage increasing from ~45% to 100%, validating the effectiveness of the SDSC method in promoting phase stability and film continuity. Although the SDSC method successfully achieved full rutile phase coverage and enhanced crystalline quality, addressing surface roughness remains a challenge requiring further improvement.

## AUTHOR DECLARATIONS

### Conflict of Interest

The authors have no conflicts to disclose.

### Author Contributions

Imteaz Rahaman: Data curation (lead); Formal analysis (lead); Investigation (lead); Methodology (lead), Writing-original draft (lead); Botong Li: Data curation(supporting); Writing – review & editing (supporting). Bobby Duersch: Data curation (supporting); Writing – review & editing (supporting), Formal analysis (supporting). Hunter D. Ellis: Writing – review & editing (supporting). Kai-Fu: Conceptualization (lead); Writing – review & editing (lead); Supervision (lead); Project administration (lead); Resources (lead).

## ACKNOWLEDGEMENT

The authors would like to express their sincere gratitude for the financial support received from the University of Utah's start-up fund and the PIVOT Energy Accelerator Grant U-7352FuEnergyAccelerator2023. Furthermore, they acknowledge the instrumental facilities provided by the University of Utah, which include the Utah Nanofab Cleanroom, the Material Characterization Meldrum, and the Nanofab Electron Microscopy and Surface Analysis Facilities.

## DATA AVAILABILITY

The data that supports the findings of this study are available from the corresponding authors upon reasonable request.

For Table of Contents Use Only

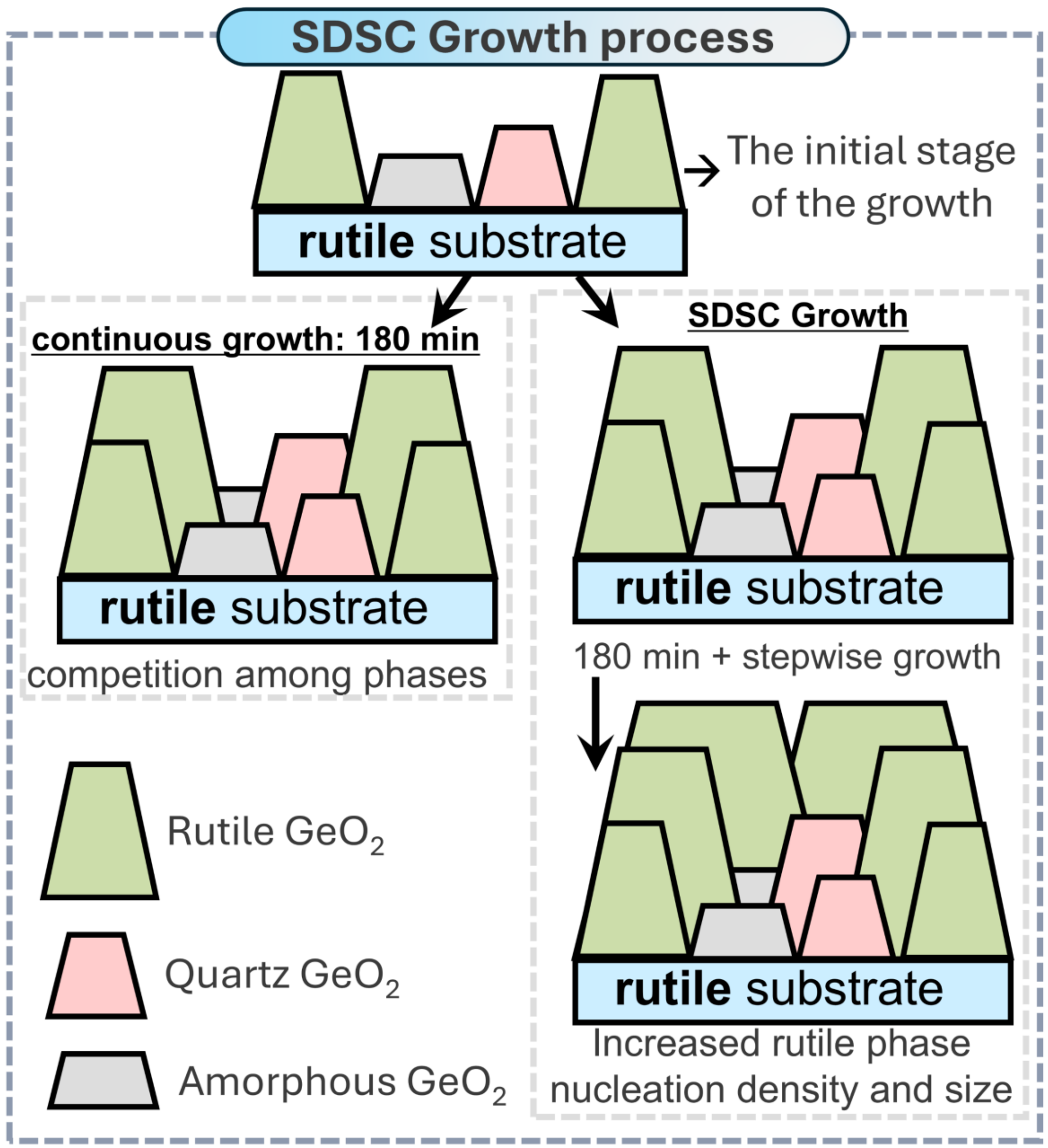